\documentclass[journal]{IEEEtran}
\IEEEoverridecommandlockouts
\usepackage{cite}
\usepackage[pdftex]{graphicx}
\usepackage{amsmath}
\usepackage{amssymb}

\usepackage{algorithmic}
\usepackage{array}
\usepackage[caption=false,font=footnotesize]{subfig}
\usepackage{dblfloatfix}
\usepackage{url}

\def\BibTeX{{\rm B\kern-.05em{\sc i\kern-.025em b}\kern-.08em T\kern-.1667em\lower.7ex\hbox{E}\kern-.125emX}}
    
\begin{document}

\title{Complementary-Aperture Pulse Sequencing for Fundamental-Band Nonlinearity Parameter Imaging}

\author{
    \IEEEauthorblockN{Esteban Avilés\textsuperscript{1,*}}
    , \IEEEauthorblockN{Michael Oelze\textsuperscript{2}}
    , \IEEEauthorblockN{Roberto Lavarello\textsuperscript{1}}
    , \IEEEauthorblockN{Andres Coila\textsuperscript{1}}
    \\
    \IEEEauthorblockA{\textsuperscript{1}\textit{Laboratorio de Imágenes Médicas, Departamento de Ingeniería, Pontificia Universidad Católica del Perú, Lima 15088, Peru}}\\
    \textsuperscript{2}\textit{Beckman Institute for Advanced Science and Technology, Department of Electrical and Computer Engineering, University of Illinois Urbana-Champaign, Urbana, Illinois 61801, USA}\\
    \textsuperscript{*}E-mail: esteban.aviles@pucp.edu.pe
}

\maketitle

\begin{abstract}
    The depletion method (DM) enables the estimation of the acoustic nonlinearity parameter, \textit{B/A}, in pulse-echo ultrasound using fundamental-band signals from low- and high-pressure acquisitions, but requires accurate knowledge of the acoustic-pressure scaling factor. This requirement limits experimental translation because of the slightly nonlinear scaling between excitation voltage set on the ultrasound system and pressure at the probe. We propose a complementary-aperture pulse sequence (CAPS) that recovers the linearly scaled low-pressure signal by coherently summing radio-frequency acquisitions from interleaved transmit subapertures, whereas a full-aperture transmission provides the high-pressure signal maintaining the same excitation voltage configured in the ultrasound system. In silico phantoms of a homogeneous sample and a reference evaluated four aperture partitions of CAPS and compared them with the DM under relative scaling factor errors from 0.01\% to 10\%. Using the DM, errors of 0.1\% and 1\% produced absolute \textit{B/A} biases of up to 27.3\% and 58.3\%, respectively. CAPS yielded mean \textit{B/A} values of 8.9 for a prescribed value of 9, with biases of less than 1.3\% across all partitions, matching an ideally calibrated DM. These results demonstrate that CAPS preserves depletion-based \textit{B/A} estimation without using the pressure scaling factor as a reconstruction input, supporting experimental and in vivo implementation without pressure-ratio calibration.
\end{abstract}


\section{Introduction}

Estimation of the acoustic nonlinearity parameter, \textit{B/A}, may aid biological tissue characterization, complementing conventional pulse-echo ultrasonography \cite{panfilova2021}. One potential clinical application is the assessment of metabolic dysfunction-associated steatotic liver disease (MASLD), because fatty liver tissue has shown higher \textit{B/A} values than healthy liver tissue \cite{zhang2001}. Lipid accumulation may contribute to this difference, since several types of adipose tissue also exhibit high \textit{B/A} values \cite{errabolu1988}.

Laboratory methods for estimating \textit{B/A} include thermodynamic, parametric-array, pump-wave, and finite-amplitude methods, but their instrumentation and acquisition requirements limit their use for clinical in vivo imaging \cite{panfilova2021,hamilton1998}. Most finite-amplitude methods estimate \textit{B/A} from the second-harmonic amplitude with a single probe. This restricts the transmit frequency to the lower portion of the probe bandwidth. Alternatively, custom assemblies of single-element transducers with different bandwidths have been used to recover the second-harmonic signal \cite{gong2004}, and the second-order ultrasound field technique, a variant of the pump-wave method, has also been proposed \cite{vanSloun2015}. However, dependence on custom probes might hinder clinical adoption.

Other pulse-echo methods estimate \textit{B/A} from fundamental-band depletion using multi-energy acquisitions \cite{nikoonahad1990,fatemi1996}. However, these methods require echoes from planar reflectors or wire targets embedded in the medium, limiting their applicability in vivo. More recently, Coila et al. proposed estimating \textit{B/A} by comparing fundamental-band RF signals acquired at low and high pressure levels, corresponding to quasi-linear and nonlinear propagation, respectively \cite{coila2025}. Unlike previous approaches, this method uses backscattered signals from randomly distributed scatterers. It is based on the fundamental-band depletion predicted by weakly nonlinear propagation theory for plane waves in attenuating fluids \cite{hamilton1998}.

Despite operating in a pulse-echo, fundamental-band framework, the depletion method (DM) requires accurate knowledge of the source-pressure scaling factor between the low- and high-pressure acquisitions. This requirement limits its potential for in vivo imaging because of the slightly nonlinear scaling between transmit voltage set on the ultrasound system and acoustic pressure at the probe \cite{coila2025}. Although the scaling factor could be measured using a needle hydrophone near the probe surface, even a 1\% error may produce a substantial error in the estimated depletion.

To address this limitation, this work introduces a complementary-aperture pulse sequence (CAPS) for fundamental-band \textit{B/A} imaging. The low-pressure signal is formed by coherently summing RF acquisitions transmitted from complementary groups of elements, whereas the high-pressure signal is acquired using the full transmit aperture. All transmissions use the same per-element excitation and the full receive aperture. By imposing the scaling through the aperture geometry, CAPS removes the need for pressure-ratio calibration while retaining compatibility with programmable arrays and randomly distributed tissue scatterers.

\section{Theory}

\subsection{\textit{B/A} Estimation from Fundamental-Band Depletion}

The DM is derived from the theory of weakly nonlinear plane-wave propagation in attenuating fluid-like media \cite{hamilton1998}. For a low-level source pressure $P_0$, the fundamental-band pressure at depth $z$ is

\begin{equation}
    P_L=P_0e^{-\alpha z}
    \left[
    1-\frac{1}{32}\Gamma^2
    \left[1-e^{-2\alpha z}\right]^2
    \right],
\end{equation}
where $\alpha$ is the attenuation coefficient and $\Gamma$ is the Gol'dberg number \cite{kinsler2000}, given by

\begin{equation}
    \Gamma=\frac{2\pi f_0\beta P_0}{\rho_0c_0^3\alpha}.
\end{equation}

Here, $f_0$ is the fundamental frequency, $\beta=1+\frac{1}{2}(B/A)$ is the Beyer nonlinearity coefficient, $\rho_0$ is the equilibrium density, and $c_0$ is the speed of sound. For a high-level source pressure $\nu P_0$, where $\nu$ is the pressure-scaling factor, the fundamental-band pressure is

\begin{equation}
    P_H=\nu P_0e^{-\alpha z}
    \left[
    1-\frac{1}{32}\nu^2\Gamma^2
    \left[1-e^{-2\alpha z}\right]^2
    \right].
\end{equation}

Nonlinear propagation transfers energy from the fundamental band to higher harmonics. Consequently, $P_H$ becomes lower than the linear prediction $\nu P_L$. In the weak-wave regime, $\Gamma<1$, the corresponding depletion in the backscattered signal is

\begin{equation}
    \nu P_L'-P_H'=
    \frac{\nu^3-\nu}{32}
    P_0\Gamma^2
    \left[1-e^{-2\alpha z}\right]^2
    \frac{\gamma}{z}e^{-2\alpha z},
\end{equation}
where $P_L'$ and $P_H'$ are the fundamental-band envelopes of the backscattered RF signals acquired at the low and high pressure levels, respectively, and $\gamma$ is the reflection coefficient.

Because $\gamma$ is generally unknown, the sample measurements are calibrated using a reference phantom with known acoustic properties \cite{coila2025}. Then, in a selected region of interest (ROI), the nonlinearity coefficient is estimated as

\begin{equation}
    \beta
    = \beta_R
    \sqrt{
    \frac{
    \left|\nu P_L'-P_H'\right|P_{L,R}'
    }{
    \left|\nu P_{L,R}'-P_{H,R}'\right|P_L'
    }}
    \frac{1-e^{-2\alpha_Rz}}{1-e^{-2\alpha z}}
    \frac{\alpha}{\alpha_R},
    \label{eq:forward_model}
\end{equation}
where the subscript $R$ denotes reference-phantom variables.

\subsection{Sensitivity to Pressure-Scaling Error}

The DM requires knowledge of the acoustic pressure scaling factor $\nu$. Because the pressure ratio cannot generally be inferred from the transmit-voltage ratio, $\nu$ must be measured, for example using a needle hydrophone \cite{coila2025}. Since depletion is the difference between two nearly equal signals, errors in $\nu$ can be strongly amplified. Let $D=\nu P_L'-P_H'$ denote the true depletion and $\widehat{\nu}=\nu + \Delta\nu$ the measured scaling factor. The estimated depletion is then $\widehat{D}=\widehat{\nu} P_L' - P_H'=D+\Delta\nu P_L'$. Using the weak-wave model, define

\begin{equation}
    q(z)=
    \frac{\Gamma^2}{32}
    \left[1-e^{-2\alpha z}\right]^2,
\end{equation}
such that

\begin{equation}
    \frac{D}{P_L'}
    =
    \frac{\nu[\nu^2-1]q(z)}
    {1-q(z)}.
\end{equation}

Therefore, defining the signed relative scaling error as $\varepsilon_\nu=\Delta\nu/\nu$, and since $q(z)\ll1$, the relative perturbation of the depletion can be approximated as

\begin{equation}
    \frac{|\widehat{D}-D|}{D}
    \approx
    \frac{32|\varepsilon_\nu|}
    {[\nu^2-1]\Gamma^2
    \left[1-e^{-2\alpha z}\right]^2}.
    \label{eq:depletion_scaling_error}
\end{equation}

\subsection{Complementary-Aperture Pulse Sequence}

The proposed CAPS partitions an $N$-element linear array into $m$ complementary interleaved subapertures, where $m$ divides $N$. Each subaperture contains $N/m$ elements, selected every $m$-th element, while an additional transmission uses the full aperture. All transmissions use the same per-element excitation and transmit delays, with the full aperture active during reception.

Under linear propagation, the pressure generated by an active element set $\mathcal{S}$ at a spatial position $\mathbf{x}$ is the superposition of its individual element contributions,

\begin{equation}
    p_{\mathcal{S}}(\mathbf{x})
    =
    \sum_{n\in\mathcal{S}}p_n(\mathbf{x}).
\end{equation}

Because the $m$ subapertures are complementary and collectively contain all array elements, their fields satisfy

\begin{equation}
p_F^{\mathrm{lin}}(\mathbf{x})=\sum_{j=1}^{m}p_j(\mathbf{x}),
\end{equation}
where $p_j$ denotes the field generated by the $j$-th subaperture and $p_F^{\mathrm{lin}}$ denotes the linearly propagated full-aperture field. Consequently, coherent summation of the corresponding RF signals reconstructs the linear full-aperture response.

For a uniformly excited linear array with $M$ active elements, the on-axis pressure in the far-field can be approximated as

\begin{equation}
    p_M(z)
    \approx
    p_e(z)
    \lim_{\psi\rightarrow0}
    \frac{\sin(M\psi/2)}{\sin(\psi/2)}
    =
    Mp_e(z),
\end{equation}
where $p_e$ is the single-element pressure and $\psi$ is the phase difference between adjacent active elements. Since each subaperture contains $N/m$ elements, its axial pressure is approximately $1/m$ of the full-aperture pressure. The sequence therefore provides a nominal pressure-scaling factor of $\nu\approx m$.

The subaperture RF signals must be coherently summed before envelope detection so that their fundamental-band envelope approximates $\nu P_L'$. The depletion can then be evaluated as

\begin{equation}
    D_{\mathrm{CAPS}}=P_{L,\Sigma}'-P_H',
\end{equation}
where $P_{L,\Sigma}'$ is obtained from the coherently summed RF signals. This formulation imposes the pressure-scaling factor through the aperture partition rather than through an independent pressure measurement. Although each sparse subaperture produces grating lobes individually, coherent summation of all complementary acquisitions reconstructs the full-aperture linear response and suppresses their combined contribution.

\section{Methods}

\subsection{Generation of Simulated RF Data}

RF data were generated using the nonlinear acoustic solver of the k-Wave toolbox \cite{treeby2012}. The two-dimensional computational domain measured $60\times40$~mm with an isotropic grid spacing of $0.02$~mm. Both media had a sound speed of $1500$~m/s, a mean density of $1000$~kg/m$^3$, and quadratic attenuation defined as $\alpha(f)=0.10f^2$~dB/cm, with $f$ expressed in MHz. Spatial density fluctuations with a standard deviation of $2\%$ of the mean were introduced to generate randomly distributed scatterers. The same density map was used for all transmissions of the same medium.

A homogeneous sample phantom with \textit{B/A} = 9 and a homogeneous reference phantom with \textit{B/A} = 6 were simulated. The linear array had 128 elements with a width and pitch of $0.3$~mm, resulting in a $38.4$ mm aperture. All elements were active during reception. Plane-wave transmission was simulated using a Gaussian tone burst centered at $5$~MHz with a $22.6\%$ fractional bandwidth at $-6$~dB.

\subsection{Numerical Evaluation}

For each partition $m=2$, $4$, $8$, and $16$, corresponding to a nominal scaling factor $\nu=m$, the CAPS acquisition comprised $m+1$ transmissions. The first $m$ transmissions separately excited the complementary interleaved subapertures at the same nominal per-element source pressure of $400$~kPa, while one full-aperture transmission at $400$~kPa provided the high-pressure signal. An additional full-aperture transmission at $400/m$~kPa was simulated to provide the DM low-pressure signal.

To isolate the effect of scaling factor error, the full-aperture RF data acquired at $400/m$ and $400$~kPa were held fixed while the true factor $\nu$ in~\eqref{eq:forward_model} was replaced by $\widehat{\nu}=\nu(1+\varepsilon_\nu)$. Relative errors of $\varepsilon_\nu=0$, $\pm0.01\%$, $\pm0.1\%$, $\pm1\%$, and $\pm10\%$ were evaluated for each value of $\nu$.

For the CAPS evaluation, the $m$ complementary-subaperture RF signals were coherently summed before envelope detection. The summed signal was used as the linearly scaled low-pressure signal in the depletion estimator, without providing $\nu$ as an input.

For both evaluations, the mean estimated \textit{B/A} was calculated within a rectangular ROI extending from $1$ to $5$~cm in depth and spanning $2.5$~cm laterally, centered at the array midline, and compared with the ground-truth value.

\section{Results}

\subsection{Sensitivity of the DM to Scaling Factor Error}

\begin{figure}[h!]
    \centering
    \includegraphics[width=\linewidth]{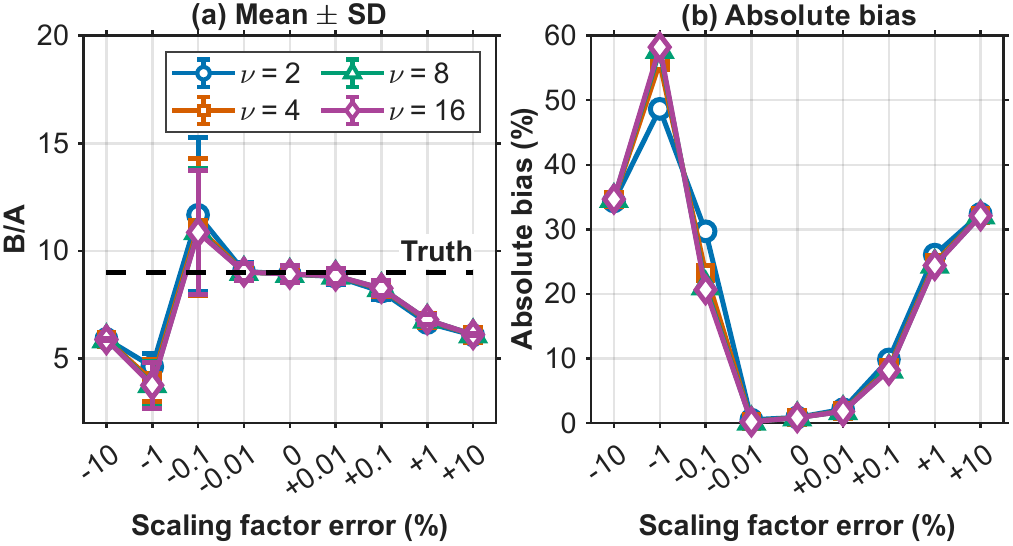}
    \caption{Effect of scaling factor error on the depletion method estimates (a) ROI mean $\pm$ spatial standard deviation (SD) of the \textit{B/A} and (b) absolute bias for $\nu=2,4,8,$ and $16$. The dashed line denotes the prescribed \textit{B/A} = 9.}
    \label{fig:scaling_error}
\end{figure}

With the correct scaling factor, the mean \textit{B/A} was 8.9 for all $\nu$, corresponding to 1.2\% bias. The bias remained below 2.4\% for $|\varepsilon_\nu|=0.01\%$, but increased to 8.4--27.3\% at 0.1\% and 24.4--58.3\% at 1\% (see Fig.~\ref{fig:scaling_error}), consistent with the proportional dependence of the depletion error on $|\varepsilon_\nu|$ in~\eqref{eq:depletion_scaling_error}. The similar responses for $\nu\geq4$ are also predicted theoretically: because $P_0=400/\nu$ kPa and $\Gamma \propto P_0$, the denominator $(\nu^2 -1)\Gamma^2$ is proportional to $1-1/\nu^2$, which changes little for larger $\nu$. The sign asymmetry follows from $\widehat{D}=D+\Delta \nu P_L'$: positive errors progressively drive the sample-to-reference depletion ratio toward unity, whereas moderate negative errors reduce the smaller reference depletion in the denominator of~\eqref{eq:forward_model} toward zero first, producing the overestimation observed at $-0.1\%$. At $|\varepsilon_\nu|=10\%$, the error term dominated both depletion terms, driving their ratio toward unity. Because the sample and reference had the same attenuation, the attenuation correction also became unity, pulling the estimated sample \textit{B/A} toward the reference value of 6. The resulting estimates of 5.88--6.11 were within 2\% of the reference, with a 32.1--34.7\% bias relative to the sample ground truth.

\begin{figure}[h!]
    \centering
    \includegraphics[width=\linewidth]{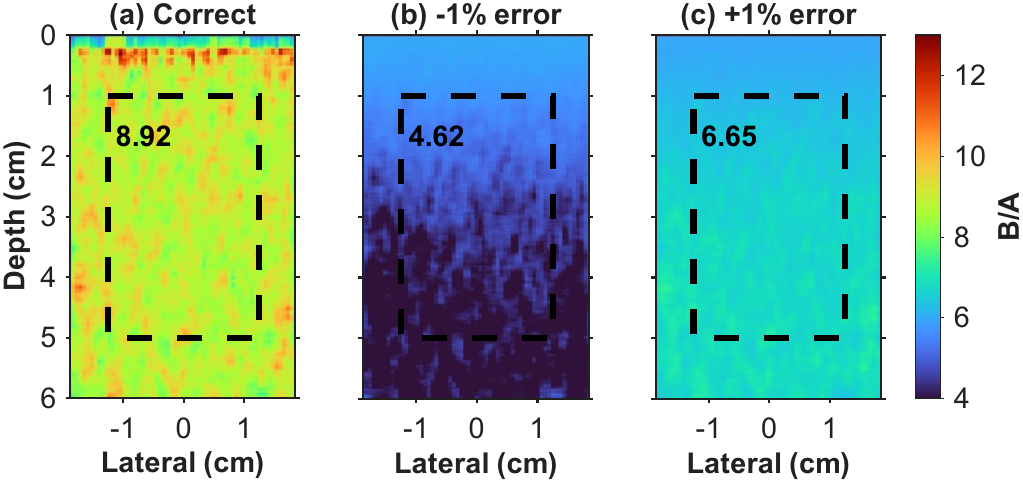}
    \caption{Representative \textit{B/A} maps with the depletion method for $\nu=2$ using (a) the correct scaling factor, (b) $-1$\% error, and (c) $+1$\% error. Dashed rectangles mark the ROI; numbers are the mean within the ROI.}
    \label{fig:example_errors}
\end{figure}

Figure~\ref{fig:example_errors} further illustrates the sign-dependent effect for the representative case $\nu=2$. With the correct scaling factor, the map remained homogeneous and the ROI \textit{B/A} was 8.9$\pm$0.4, close to the configured value of 9. A $-1$\% error reduced the estimate to 4.6 (48.8\% bias), whereas $+1$\% produced 6.7 (26.1\% bias).

\subsection{\textit{B/A} Estimation using CAPS}

\begin{figure}[h!]
    \centering
    \includegraphics[width=0.75\linewidth]{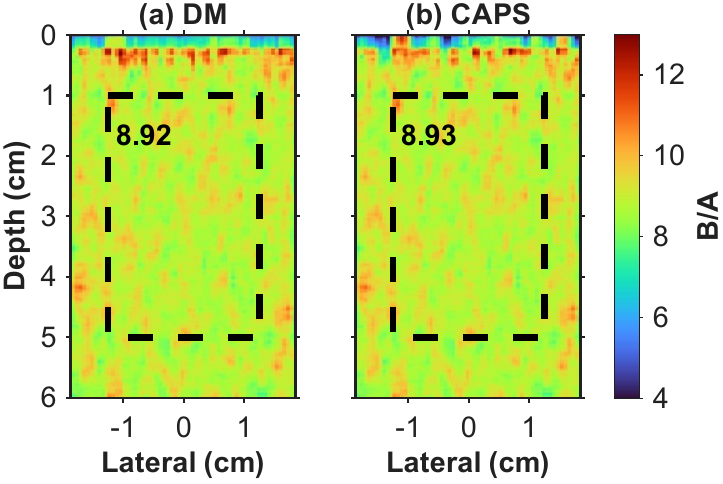}
    \caption{Representative \textit{B/A} maps for $\nu=2$ obtained with (a) the depletion method using the correct scaling factor and (b) CAPS. Dashed rectangles mark the ROI; numbers are the ROI means.}
    \label{fig:cps_example}
\end{figure}

Figure~\ref{fig:cps_example} compares representative \textit{B/A} maps obtained with the correctly calibrated DM and CAPS for $\nu=2$. Both methods produced a mean of 8.9, corresponding to 1.2\% bias. CAPS remained consistent for $m=2,4,8,$ and $16$, with the same mean and bias, despite the increasingly sparse subapertures. This stability is consistent with coherent summation suppressing their grating-lobe contributions and reconstructing the full-aperture linear response.

\section{Discussion and Conclusion}

This work introduced CAPS for fundamental-band \textit{B/A} imaging without providing the pressure-scaling factor to the estimator. The sensitivity analysis identifies the pressure scaling factor uncertainty as a critical failure mode of depletion-based \textit{B/A} imaging. Errors of only 0.1\% produced biases of up to 27.3\%, while 1\% errors produced biases of up to 58.3\%, with a strong dependence on error sign. At larger errors, the estimates approached the reference \textit{B/A} because both phantoms shared the same attenuation coefficient. The error amplification agrees with the theoretical prediction from~\eqref{eq:depletion_scaling_error}, whereas the sign dependence and convergence toward the reference value follow from the exact relation $\widehat{D} = D+\Delta \nu P_L'$ applied to the sample and reference depletion signals.

CAPS constructs the required linearly scaled low-pressure signal without using $\nu$ as a reconstruction input and reproduced the correctly calibrated DM estimates for all tested aperture partitions. Relative to \cite{coila2025}, it removes the need for independent pressure-ratio calibration while retaining fundamental-band operation and compatibility with randomly distributed scatterers. Unlike reflector-based approaches \cite{nikoonahad1990,fatemi1996}, CAPS remains applicable to tissue echoes acquired with programmable arrays.

This study was limited to two-dimensional simulations with idealized array operation. Under the phase-coherent and aberration-free conditions considered, coherent summation suppressed the grating lobes generated by the individual sparse subapertures, explaining the stable estimates obtained even for $m=16$. Electronic noise, motion, element mismatch, and phase aberration could prevent this cancellation from being exact, particularly for larger $m$, which require sparser subapertures and more transmissions. Future work should therefore evaluate CAPS in heterogeneous and aberrating media and validate it in physical phantoms. CAPS could also be incorporated into depletion-based spatial compounding \cite{aviles2024, aviles2025} and joint attenuation and \textit{B/A} estimation \cite{merino2025}. In addition, CAPS can be coupled with closed-form weak-wave expressions for arbitrary power-law attenuation \cite{aviles2026} to extend it to biological tissues, whose attenuation is closer to linear in frequency.

In conclusion, CAPS can provide \textit{B/A} maps equivalent to those of the correctly calibrated depletion method and stable estimates across several aperture partitions. By replacing a measured pressure scaling ratio with coherent aperture summation, CAPS may remove a major calibration barrier to the experimental and in vivo translation of depletion-based \textit{B/A} imaging.

\section*{Acknowledgment}

This research was funded by the Consejo Nacional de Ciencia, Tecnología e Innovación Tecnológica (CONCYTEC) and the Programa Nacional de Investigación Científica y Estudios Avanzados (PROCIENCIA) under the contest E073-2025-01 ``Tesis de Pregrado y Postgrado en Ciencia, Tecnología e Innovación Tecnológica" (award number PE501099673-2025).

\ifCLASSOPTIONcaptionsoff
  \newpage
\fi


\begin{thebibliography}{00}
    \bibitem{panfilova2021} A. Panfilova, R. J. G. van Sloun, H. Wijkstra, O. A. Sapozhnikov, and M. Mischi, ``A review on \textit{B/A} measurement methods with a clinical perspective,'' \textit{J. Acoust. Soc. Am.}, vol. 149, no. 4, pp. 2200--2237, Apr. 2021, doi: 10.1121/10.0003627.
    \bibitem{zhang2001} D. Zhang, X.-F. Gong, and X. Chen, ``Experimental imaging of the acoustic nonlinearity parameter \textit{B/A} for biological tissues via a parametric array,'' \textit{Ultrasound Med. Biol.}, vol. 27, no. 10, pp. 1359--1365, Oct. 2001, doi: 10.1016/S0301-5629(01)00432-X.
    \bibitem{errabolu1988} R. L. Errabolu, C. M. Sehgal, R. C. Bahn, and J. F. Greenleaf, ``Measurement of ultrasonic nonlinear parameter in excised fat tissues,'' \textit{Ultrasound Med. Biol.}, vol. 14, no. 2, pp. 137--146, Mar. 1988, doi: 10.1016/0301-5629(88)90181-0.
    \bibitem{hamilton1998} M. F. Hamilton and D. T. Blackstock, \textit{Nonlinear Acoustics}. San Diego, CA, USA: Academic Press, 1998.
    \bibitem{gong2004} X. Gong, D. Zhang, J. Liu, H. Wang, Y. Yan, and X. Xu, ``Study of acoustic nonlinearity parameter imaging methods in reflection mode for biological tissues,'' \textit{J. Acoust. Soc. Am.}, vol. 116, no. 3, pp. 1819--1825, Sep. 2004, doi: 10.1121/1.1781709.
    \bibitem{vanSloun2015} R. J. G. van Sloun, L. Demi, C. Shan, and M. Mischi, ``Ultrasound coefficient of nonlinearity imaging,'' \textit{IEEE Trans. Ultrason., Ferroelectr., Freq. Control}, vol. 62, no. 7, pp. 1331--1341, Jul. 2015, doi: 10.1109/TUFFC.2015.007009.
    \bibitem{nikoonahad1990} M. Nikoonahad and D. C. Liu, ``Pulse-echo single frequency acoustic nonlinearity parameter (\textit{B/A}) measurement,'' \textit{IEEE Trans. Ultrason., Ferroelectr., Freq. Control}, vol. 37, no. 3, pp. 127--134, May 1990, doi: 10.1109/58.55301.
    \bibitem{fatemi1996} M. Fatemi and J. F. Greenleaf, ``Real-time assessment of the parameter of nonlinearity in tissue using `nonlinear shadowing','' \textit{Ultrasound Med. Biol.}, vol. 22, no. 9, pp. 1215--1228, 1996, doi: 10.1016/S0301-5629(96)00140-8.
    \bibitem{coila2025} A. Coila, A. Romero, M. L. Oelze, and R. Lavarello, ``Nonlinearity parameter estimation method from fundamental band signal depletion in pulse-echo using a dual-energy model,'' \textit{J. Acoust. Soc. Am.}, vol. 157, no. 3, pp. 1969--1980, Mar. 2025, doi: 10.1121/10.0036215.
    \bibitem{kinsler2000} L. E. Kinsler, A. R. Frey, A. B. Coppens, and J. V. Sanders, \textit{Fundamentals of Acoustics}, 4th ed. New York, NY, USA: Wiley, 2000.
    \bibitem{treeby2012} B. E. Treeby, J. Jaros, A. P. Rendell, and B. T. Cox, ``Modeling nonlinear ultrasound propagation in heterogeneous media with power law absorption using a $k$-space pseudospectral method,'' \textit{J. Acoust. Soc. Am.}, vol. 131, no. 6, pp. 4324--4336, Jun. 2012, doi: 10.1121/1.4712021.
    \bibitem{aviles2024} E. Avil\'es, R. Lavarello, and A. Coila, ``Nonlinearity parameter imaging of local estimates using spatial compounding,'' in \textit{Proc. IEEE Ultrason., Ferroelectr., Freq. Control Joint Symp. (UFFC-JS)}, Taipei, Taiwan, Sep. 2024, pp. 1--4, doi: 10.1109/UFFC-JS60046.2024.10793740.
    \bibitem{aviles2025} E. Avil\'es, R. Lavarello, and A. Coila, ``Nonlinearity parameter imaging using a multi-view joint inverse problem formulation,'' in \textit{Proc. IEEE Int. Ultrason. Symp. (IUS)}, Utrecht, The Netherlands, Sep. 2025, pp. 1--4, doi: 10.1109/IUS62464.2025.11201336.
    \bibitem{merino2025} S. Merino, A. Romero, R. Lavarello, and A. Coila, ``Regularized joint estimator of the nonlinearity parameter and attenuation coefficient using a nonlinear least-squares algorithm,'' \textit{Ultrason. Imag.}, vol. 47, no. 6, pp. 270--282, Nov. 2025, doi: 10.1177/01617346251362389.
    \bibitem{aviles2026} E. Avil\'es, M. Oelze, R. Lavarello, and A. Coila, ``Closed-form weak-wave propagation pressure amplitudes for arbitrary power-law attenuation,'' 2026, \textit{arXiv:2609.19458}, doi: 10.48550/arXiv.2609.19458.

\end{thebibliography}
\end{document}